\documentclass[unnumsec,webpdf,contemporary,large]{oup-authoring-template}%

\graphicspath{{Fig/}}

\theoremstyle{thmstyleone}%
\theoremstyle{thmstyletwo}%
\theoremstyle{thmstylethree}%

\usepackage{arydshln}
\usepackage{hyperref}
\usepackage{subfigure}
\usepackage{subcaption}

\begin{document}

\journaltitle{Journal Title Here}
\DOI{DOI HERE}
\copyrightyear{2022}
\pubyear{2019}
\access{Advance Access Publication Date: Day Month Year}
\appnotes{Paper}

\firstpage{1}


\title[]{Predicting Biomedical Interactions with Bayesian inference over Graph Convolutional Network Structures}

\author[1]{Kishan KC}
\author[2]{Paribesh Regmi}
\author[2,$\ast$]{Rui Li}
\author[3]{Anne R. Haake}

\authormark{KC et al.}

\address[1]{\orgname{Amazon}, \postcode{94089}, \state{California}, \country{USA}}
\address[2]{\orgdiv{GCCIS}, \orgname{RIT}, \orgaddress{\street{1 Lomb Memorial Dr.}, \postcode{14623-5603}, \state{New York}, \country{USA}}}
\corresp[$\ast$]{Corresponding author. \href{email:email-id.com}{rxlics@rit.edu}}

\received{Date}{0}{Year}
\revised{Date}{0}{Year}
\accepted{Date}{0}{Year}



\abstract{\textbf{Motivation}\\
Heterogeneous molecular entities and their interactions, commonly depicted as a network, are crucial for advancing our systems-level understanding of biology. With recent advancements in high-throughput data generation and a significant improvement in computational power, graph neural networks (GNNs) have demonstrated their effectiveness in predicting biomedical interactions. Since GNNs follow a neighborhood aggregation scheme, the number of graph convolution (GC) layers (i.e., depth) determines the neighborhood orders from which they can aggregate information, thereby significantly impacting the model's performance. However, it often relies on heuristics or extensive experimentation to determine an appropriate GNN depth for a given biomedical network. These methods can be unreliable or result in expensive computational overhead. Moreover, GNNs with more GC layers tend to exhibit poor calibration, leading to high confidence in incorrect predictions.\\
\textbf{Results}\\
To address these challenges, we propose a Bayesian model selection framework to jointly infer the most plausible number of GC layers supported by the data, apply dropout regularization, and learn network parameters. Experiments on four biomedical interaction datasets demonstrate that our method achieves superior performance over competing methods, providing well-calibrated predictions by allowing GNNs to adapt their depths to accommodate interaction information from various biomedical networks..\\
\textbf{Availability}\\
Source code and data is available at \url{https://github.com/kckishan/BBGCN-LP/tree/master}
}


\maketitle

\section{Introduction}
Biomedical interaction networks represent complex interplay between heterogeneous molecular entities in a system, such as protein-protein interactions (PPIs) \cite{luck2020reference},
drug-drug interactions (DDIs) \cite{zitnik2018modeling}, drug-target interactions (DTIs) \cite{luo2017network} and gene-disease associations (GDIs) \cite{agrawal2018large}. The study of interaction networks advances our understanding of system-level understanding of biology, and sheds light on the molecular mechanisms of a given phenotype \cite{cowen2017network}. Despite the continuous growth of omics data driven by advancements in high-throughput technologies, existing biological networks remain characterized by noise, sparsity, and incompleteness. This is primarily due to the ongoing challenges in identifying biologically significant interactions that are previously unmapped.

A wide spectrum of computational methods are developed to predict novel biomedical interactions. Network-based approaches exploit available interactions to predict missing ones \cite{Cheng2012:Prediction, Wu2016:Silico, Wu2017:Sdtnbi}. They measure similarity of network properties between biological entities as an indication of interaction. Specifically, the Triadic Closure Principle (TCP) suggests that biological entities with shared neighbors are likely to interact with each other. However, this common neighbor hypothesis fails to identify interactions between most protein pairs in PPI prediction. To solve the problem, L3 heuristic \cite{kovacs2019network} further assumes that biological entities linked by multiple paths of length 3 are more likely to have direct links/interactions.

Network embedding approaches \cite{perozzi2014deepwalk,grover2016node2vec} aim to transform network nodes into low-dimensional vector representations while preserving key network properties. Deepwalk \cite{perozzi2014deepwalk} achieves this by generating truncated random walks of length $l$ and defining a context window of size $k$ to represent the neighborhood of each node. It learns distributed representations for network nodes based on these random walks. However, Deepwalk's approach relies on a rigid notion of network neighborhood, making it insensitive to the unique connectivity patterns present in a network. On the other hand, node2vec \cite{grover2016node2vec} employs a more flexible strategy by using biased random walks that strike a balance between breadth-first and depth-first search. These network embedding methods generate feature representations optimized to preserve node neighborhoods in a low-dimensional feature space. Subsequently, these embeddings are utilized to train downstream classifiers for making predictions. It's worth noting that these methods primarily capture the structural information of biological networks and do not effectively integrate node-specific features.

Recent deep learning approaches for network datasets achieve great success across various domains, such as social networks \cite{Li2019:Encoding, gnn@social}, recommendation systems \cite{gnn@recommendation}, chemistry \cite{gilmer2017neural}, citation networks \cite{kipf2016variational}. In particular, graph convolutional networks (GCNs) show promising performance in biomedical interaction prediction \cite{huang2020skipgnn,kc2021predicting}. Since GCN-based methods follow a message-passing mechanism that recursively aggregates information from its neighbors, the representation of a node generated by a GCN model with $L$ graph convolution (GC) layers can capture topological information from its $L$-hop neighborhood. Although higher-order neighborhood information are important for the incomplete and sparse biomedical networks \cite{kc2021predicting}, deep GCNs with large $L$ suffer from over-smoothing \cite{Li2018:Deeper}, leading to a decline in performance as the number of GC layers increases \cite{kipf2016semi}. Consequently, determining the appropriate depth (i.e., the number of layers) for a GCN to effectively model biomedical interaction networks while avoiding over-smoothing has emerged as a significant challenge.

To address the challenge, we propose a Bayesian model selection framework to jointly infer the depth of GCN models warranted by data and perform dropout regularization \cite{kc2021joint}. To enable the number of GC layers in the encoder to go to infinity in theory, we model the depth of the GCN model as a stochastic process by defining a beta process over the number of GC layers. The beta process induces layer-wise activation probabilities that modulate neuron activations via a conjugate Bernoulli process. We specify a bilinear decoder to reconstruct interactions based on feature representations produced by the GCN model. The encoder-decoder framework constructs an end-to-end trainable model for interaction prediction.

We evaluate our inference framework by comparing it with competing GCN-based models for biomedical interaction prediction, demonstrating that our framework effectively addresses overfitting and over-smoothing issues by inferring the most appropriate depth. It strikes a balance between the GCN model's depth and neuron activations. The experiments suggest that our approach outperforms others, particularly on sparse interaction networks as it allows the GCN-based encoder's structure to dynamically adapt to accommodate incremental data. Furthermore, we find that our approach also yields well-calibrated predictions.

In summary, our contributions are as follows:
\begin{itemize}
    \item We propose a Bayesian model selection framework to jointly infer the most plausible depth and neuron activations for GCN-based encoders to learn representation for biomedical entities.
    \item Our experiments demonstrate that the framework achieves superior performance by dynamically balancing GCN depth and width.
    \item Our method enables the structure of the encoders to dynamically evolve to accommodate incrementally available interactions.
\end{itemize}

\section{Materials and Methods}
\subsection*{Biomedical interaction prediction}
A biomedical network is a network with biomedical entities as nodes and their interactions as edges. Formally, a biomedical network can be defined as $\mathcal{G} = (\mathcal{V}, \mathcal{E}, \mathbf{X})$ where $\mathcal{V}$ denotes the set of nodes representing biomedical entities such as proteins, genes, drugs, diseases and $\mathcal{E}$ represents the set of interaction between these entities. We denote the network $\mathcal{G}$ with a binary adjacency matrix $\mathbf{A}$ where $A_{ij} \in \{0, 1\}^{|\mathcal{V}| \times |\mathcal{V}|}$ represents the presence or absence of an edge between nodes. In particular, $A_{ij} = 1$ represents the existence of an interaction supported by experimental evidence whereas $A_{ij} = 0$ represents the absence of interaction.

\textbf{Biomedical interaction prediction}: Given a biomedical interaction network $\mathcal{G} = (\mathcal{V}, \mathcal{E}, \mathbf{X})$, we aim to learn a model $f$ to predict the probability of interaction $e_{ij}$ between nodes in the network: $f: e_{ij} \rightarrow [0, 1]$.

\subsection*{Graph convolutional encoder-decoder framework}
In this work, we follow the encoder-decoder framework~\cite{hamilton2017encdec} to learn the node representation of biological networks.  First, an encoder model maps each node in the biological network into a low-dimensional representation. Next, a decoder model reconstructs the structural information about the biological network from the learned representations. 

Formally, an encoder is a function that maps the biological entities $v \in \mathcal{V}$ to low-dimensional vector representation $\mathbf{q} \in \mathbb{R}^M$ i.e. $\text{ENC}: \mathcal{V} \rightarrow \mathbb{R}^M$. For biomedical interaction networks, GCN is a predominant choice as an encoder.

A GCN is composed of multiple GC layers. A GC layer~\cite{kipf2016semi} generates representation for nodes in the network by repeatedly aggregating information from immediate neighbors. A GC layer can be defined as:
\begin{align}
    \mathbf{H}^l = \sigma(\widehat{\mathbf{A}}\mathbf{H}^{l-1}\mathbf{W}^l)
\end{align}
where $\widehat{\mathbf{A}}$ is a symmetrically normalized adjacency matrix with self-connections $\widehat{\mathbf{A}} = \mathbf{D}^{-\frac{1}{2}} (\mathbf{A} + \mathbf{I}_{|\mathcal{V}|})\mathbf{D}^{-\frac{1}{2}}$. Let $\mathbf{W}^l$ is a trainable weight matrix for layer $l$, $\mathbf{H}^{l-1}$ and $\mathbf{H}^l$ are the input and output activations respectively. We can then define a GCN model with $L$ layers as:
\begin{align}
 \mathbf{H}^{l} = \begin{cases}
\mathbf{X} & \text{if $l = 0$}\\
\sigma (\widehat{\mathbf{A}}\mathbf{H}^{l-1}\mathbf{W}^{l}) &\text{if $l \in [1, \ldots, L]$}
\end{cases}
\end{align} 
$\mathbf{H}^L \in \mathbb{R}^{|\mathcal{V}| \times M}$ represents the representation for each entity in the network. We denote this representation as $\mathbf{H}^{out}$. Stacking $L$ layers of GC layer aggregates the network topological information from $L$-hop neighborhood. 

The decoder uses the representation generated by the encoder and reconstructs the structural properties of the biological network. For interaction prediction, a pairwise decoder is defined to predict the relationship or similarity between pairs of nodes i.e. $\text{DEC} : \mathbb{R}^M \times \mathbb{R}^M \rightarrow \mathbb{R}^+$

\subsection*{Probabilistic model selection for graph convolution-based encoder}
\begin{figure*}[!htb]
    \centering
    \includegraphics[width=\textwidth]{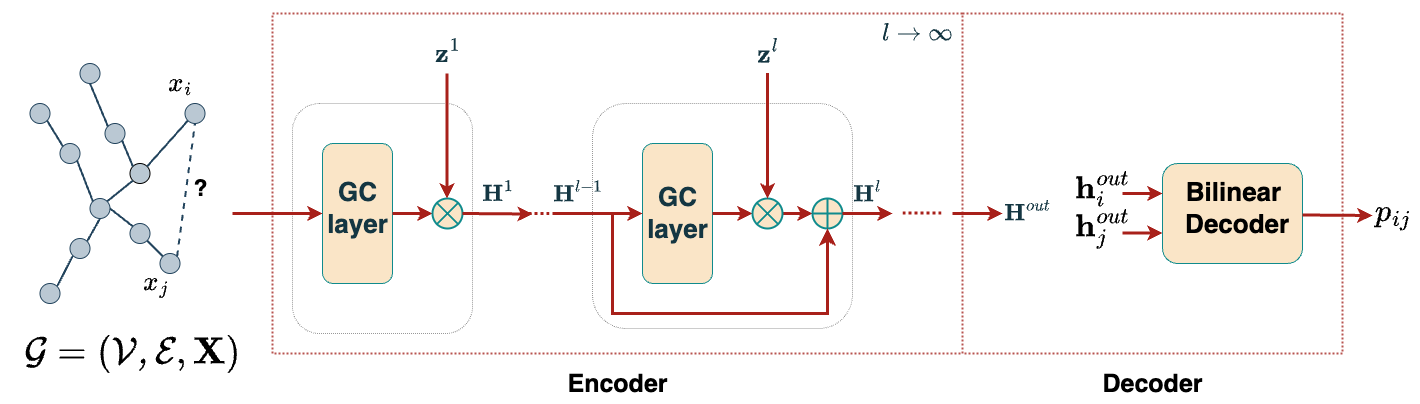}
    \caption{The schematic diagram of our proposed model features the potential for an infinite number of hidden layers within the encoder. The network takes as input a biomedical interaction network $\mathcal{G}$ with edges $\mathcal{E}$ connecting entities $\mathcal{V}$ with the feature matrix $\mathbf{X}$. The encoder is responsible for learning the representation of each entity denoted as $\mathbf{H}^{out}$. A bilinear decoder utilizes these representations $(\mathbf{h}^{out}_i, \mathbf{h}^{out}_j)$ to output a probability $p_{ij}$ of the interaction between the entities $(v_i, v_j)$.}
    \label{fig:block_diagram}
\end{figure*}

In this study, we introduce a probabilistic model selection approach~\cite{kc2021joint} to infer the optimal depth and neuron activations for the graph convolutional encoder, which is responsible for learning representations for each entity in the network. Furthermore, we utilize a bilinear decoder to reconstruct edges in the interaction network~\cite{kc2021predicting} by predicting the existence of a link (an interaction) between two entities. The schematic representation of the model is illustrated in Figure~\ref{fig:block_diagram}.
\begin{itemize}
    \item \textbf{Encoder}: a graph convolutional (GC) encoder with potentially infinite hidden layers that takes an interaction graph $\mathcal{G}$ and generates representation for each entity in the interaction network. In particular, we propose to model the depth of the graph convolution encoder as a stochastic process and jointly perform dropout regularization upon the inferred hidden layers.
    \item \textbf{Decoder}: a bilinear decoder that takes the representation of two nodes $v_i$ and $v_j$ and compute the probability $p_{ij}$ of their interaction $e_{ij}$.
\end{itemize}

We next discuss the details of each component of the proposed framework.

\subsubsection*{A GC encoder with infinite GC layers}

The issue of overfitting and over-smoothing in deep GC encoders is notable, with 2-layer models outperforming deeper ones~\cite{kipf2016semi}. On the other hand, shallow encoders struggle to capture broader neighborhoods for message aggregation. Thus, selecting the appropriate depth ($L$) for the GC encoder is a crucial step to mitigate these issues. To address this challenge, we propose to infer the most plausible depth based on the input network's characteristics by treating it as a stochastic process. Specifically, we define a beta process over the number of hidden layers~\cite{Paisley2010:stick_BP,Broderick2012:BP}. The layer activations induced by the beta process is used to modulate neuron activations in a layer by defining a conjugate Bernoulli process. We adopt a stick-breaking construction of the beta process and its conjugate Bernoulli process as:
\begin{align}\label{beta_Ber}
    z^{ml} \sim \text{Bernoulli}(\pi^l) , \quad \pi^l = \prod_{j=1}^l \nu^j, \quad \nu^l \sim \text{Beta}(\alpha, \beta)
\end{align}
where $\nu_l$ is drawn sequentially from the beta distribution, and $\pi_l$ denotes the activation probability, which decreases as the layer index $l$ increases. $z^{ml}$ represents a Bernoulli variable, with a probability of $z^{ml}=1$ equal to $\pi_l$. The value $z^{ml}=1$ signifies that the $m$-th neuron in layer $l$ is activated. With this construction, the GC encoder takes the form:
\begin{align}\label{encoder_form}
    \mathbf{H}^l = \sigma(\widehat{\mathbf{A}}\mathbf{H}^{l-1}\mathbf{W}^l) \bigotimes \mathbf{z}^l + \mathbf{H}^{l-1}, \quad l \in \{1, 2, \ldots \infty\}
\end{align}
where $\mathbf{W}^l \in \mathbb{R}^{M \times M}$ is the weight matrix of layer $l$. We regularize the output of layer $l$ by multiplying it elementwisely by a binary vector $\mathbf{z}^l$ where its element $z^{ml} \in \{0, 1\}$. Figure~\ref{fig:block_diagram} shows the network structure with binary vector $\mathbf{z}^l$ applied to the output of layer $l$. To facilitate the implementation of an infinite number of layers, we utilize skip connections between layers. These connections propagate the output from the last layer through an infinite sequence of layers, ultimately reaching the output layer \cite{kc2021joint,Regmi2023:Adavae, Regmi2025:BNA-GNN, thapa2024bayesian}.

\subsubsection*{A bilinear interaction decoder}
We adopt a bilinear decoder~\cite{kc2021predicting} to predict the probability of interactions between biomedical entities based on the latent representations of the graph obtained from the encoder and. Specifically, we define a bilinear layer that maps the representation of entities $\mathbf{h}^{out}_i$ and $\mathbf{h}^{out}_j$ to their edge representation $\mathbf{e}_{ij} \in \mathbb{R}^{d^* \times 1}$ as:
\begin{equation}
    \mathbf{e}_{ij} = \text{ELU}(\mathbf{h}^{out}_i \mathbf{W}^b \mathbf{h}^{out}_j  + \mathbf{b})
\end{equation}
where $\mathbf{e}_{ij}$ denotes the representation of edge between entities $v_i$ and $v_j$, $\mathbf{W}_b \in \mathbb{R}^{d^* \times M \times M}$ represents the learnable fusion matrix, and $\mathbf{b}$ denotes the bias of bilinear layer. The probability $p_{ij}$ of interaction between entities $v_i$ and $v_j$ is obtained by passing the edge representation $\mathbf{e}_{ij}$ through the fully-connected (FC) layer.
\begin{equation}
    p_{ij} = \text{sigmoid}(\text{FC}_2(\text{ELU}(\text{FC}_1(\mathbf{e}_{ij})))
\end{equation}

\subsection*{Efficient variational inference}
Given an interaction dataset $D = \{\mathbf{A}, \mathbf{X}\}$ with adjacency matrix $\mathbf{A} \in \mathbb{R}^{|\mathcal{V}| \times |\mathcal{V}|}$ and feature matrix $\mathbf{X}$, we aim to reconstruct the edges in the input network. For the problem of classifying the edges, we specify the likelihood of the neural network as:
\begin{equation}\label{likelihood}
    p(D|\mathbf{Z}, \mathbf{W}) = \prod_{n=1}^N \text{Bernoulli}(A_{ij}| f(\mathbf{A}, \mathbf{X}; \mathbf{Z}, \mathbf{W}))
\end{equation}
where $f(\cdot)$ represents the softmax function, $\mathbf{Z}$ is a binary matrix that represents the network structure for GC encoder whose $l$-th column is $\mathbf{z}^l$, and $\mathbf{W}$ denotes the set of weight matrices.

We define a prior over network structure $\mathbf{Z}$ based on the beta process and its conjugate Bernoulli process in Equation \ref{beta_Ber} as 
\begin{align} \label{beta_prior}
    p(\mathbf{Z}, \mathbf{\nu}| \alpha, \beta) &= p(\mathbf{\nu}|\alpha, \beta) p(\mathbf{Z}|\mathbf{\nu}) \nonumber \\
    &= \prod_{l=1}^{\infty}\text{Beta}(\nu_l|\alpha, \beta) \text{Bernoulli}(z^{ml}|\pi_l)
\end{align}

The marginal likelihood obtained by combining beta process prior in Equation~\ref{beta_prior} and the likelihood in Equation~\ref{likelihood} is:
\begin{equation}
    p(D|\mathbf{W}, L, \alpha, \beta) = \int p(D|\mathbf{W}, \mathbf{Z}) p(\mathbf{Z}, \nu|\alpha, \beta) d\mathbf{Z}d\mathbf{\nu}
\end{equation}
The precise computation of the marginal likelihood is intractable due to the neural network's non-linearity and the infinite dimensionality of $\mathbf{Z}$, which results from the infinite number of hidden layers.

To address this, we employ a structured stochastic variational inference framework introduced by~\cite{Hoffman2013:SVI,Hoffman2015:SSVI} to approximate the marginal likelihood. The lower bound for the logarithm of the marginal likelihood is expressed as follows:
\begin{align}\label{elbo}
    \log p(D|\mathbf{W}, L, \alpha, \beta) \geq \mathbb{E}&_{q(\mathbf{Z}, \nu)]}[\log p(D|\mathbf{W}, \mathbf{Z})] \nonumber \\- &\text{KL}[q(\nu)||p(\nu)] \nonumber \\ 
    - &\text{KL}[q(\mathbf{Z}|\nu)||p(\mathbf{Z}|\nu)]
\end{align}
where $q(\nu)$ and $q(\mathbf{Z}|\nu)$ represent variational distributions with a truncation level $T$. We specify them as a beta distribution and a concrete Bernoulli distribution, respectively, as:
\begin{align}
&q(\mathbf{Z}, \boldsymbol \nu|\{a^t\}_{t=1}^{T}, \{b^t\}_{t=1}^{T}) = q(\boldsymbol \nu)q(\mathbf{Z}|\boldsymbol \nu) \\= &\prod_{t=1}^{T}\text{Beta}(\nu^t|a^t, b^t)\prod_{m=1}^M \text{ConBer}(z^{mt}|\pi^t)
\label{eq:variational}
\end{align}
 where the concrete Bernoulli distribution $\text{ConBer}(z^{mt}|\pi_t)$  \cite{Maddison2016:con_dist, Jang2016:cat_rep} is a continuous relaxation of the binary variables,
permitting optimization through gradient descent.

\section*{Results and discussion}
We evaluate the performance of our method by applying it to infer the network structures of GC encoders for biomedical interaction prediction. We examine the behaviors of our proposed framework and conduct comparisons with heuristically designed encoder structures. We also demonstrate that the setting of truncation ($K$) has no impact on the performance of our model.

\subsection*{Datasets}
For experimental evaluation, we consider four publicly-available biological network datasets: (a) \textbf{BioSNAP-DTI}: Drug Target Interaction network with 15,139 drug-target interactions between 5,018 drugs and 2,325 proteins, (b) \textbf{BioSNAP-DDI}: Drug-Drug Interactions with 48,514 drug-drug interactions between 1,514 drugs extracted from drug labels and biomedical literature, (c) \textbf{HuRI-PPI}: HI-III human PPI network contains 5,604 proteins and 23,322 interactions generated by multiple orthogonal high-throughput yeast two-hybrid screens. and (d) \textbf{DisGeNET-GDI}: gene-disease network with 81,746 interactions between 9,413 genes and 10,370 diseases curated from GWAS studies, animal models, and scientific literature. For all interaction datasets, we represent the interactions as binary adjacency matrix $\mathbf{A}$ with $0$ representing the absence of an interaction and $1$ representing the presence of an interaction between two entities in a network. We perform a random split of interactions within a biological network, partitioning them into train/validation/test sets following a ratio of $7:1:2$. For each set, we acquire pairs of nodes that exhibit interactions between them (positive pairs) from the adjacency matrix $\mathbf{A}$. In order to form a balanced dataset, we randomly select an equal number of node pairs with no interactions (negative pairs).

\subsection*{Baseline Methods}
We compare our proposed method with the following baselines for interaction prediction. 
\begin{itemize}
    \item \textbf{DeepWalk}~\cite{perozzi2014deepwalk} performs truncated random walk exploring the network neighborhood of nodes and applies the skip-gram model to learn the $M$-dimensional embedding for each node in the network. Node features are concatenated to form edge representation and a logistic regression classifier is trained.
    \item \textbf{node2vec}~\cite{grover2016node2vec} extends DeepWalk by running biased random walk based on breadth/depth-first search to capture both local and global network structure.
    \item \textbf{L3}~\cite{kovacs2019network} counts the number of length 3 paths normalized by the degree for all the node pairs.
    \item \textbf{GCN}~\cite{kipf2016semi} uses normalized adjacency matrix to learn node representations. The representation for nodes are concatenated to form feature representation for the edges and the fully connected layer use these edge representation to reconstruct edges.
\end{itemize}
\subsection*{Experimental Setup}
We evaluate the baselines and our method based on how well they predict the missing interactions in a given interaction network. For the methods, we fine-tune the hyperparameters using the validation set. To gauge and compare performance, we utilize two key metrics: (a) the area under the precision-recall curve (AUPRC) and (b) the area under the receiver operating characteristics curve (AUROC). Higher AUROC and AUPRC values signify better predictive performance. 

\subsection*{Biomedical interaction prediction}

\begin{table}[htb]
\caption{Average AUPRC and AUROC with $\pm$ one standard deviation on biomedical interaction prediction}
\label{table_results}
    \centering
    \renewcommand{\arraystretch}{1.2}
    \begin{tabular}{c|ccc}
    \hline
Dataset & Method & AUPRC & AUROC \\
\hline
\multirow{5}{*}{DTI} & DeepWalk & 0.753 $\pm$ 0.008 & 0.735 $\pm$ 0.009 \\
& node2vec 		& 0.771 $\pm$ 0.005 	& 0.720 $\pm$ 0.010  \\
& L3			& 0.891 $\pm$ 0.004 	& 0.793 $\pm$ 0.006 \\
& VGAE			& 0.853 $\pm$ 0.010 	& 0.800 $\pm$ 0.010 \\
& GCN			& 0.896 $\pm$ 0.006 	& 0.914 $\pm$ 0.005 \\
\cdashline{2-4}
& Ours & \textbf{0.925  $\pm$ 0.002} & \textbf{0.933 $\pm$ 0.002} \\
\hline
\multirow{5}{*}{DDI} & DeepWalk 	& 0.698 $\pm$ 0.012 & 0.712 $\pm$ 0.009 \\
& node2vec 	& 0.801 $\pm$ 0.004 & 0.809 $\pm$ 0.002 \\ 
& L3			& 0.860 $\pm$ 0.004 & 0.869 $\pm$ 0.003 \\ 
& VGAE		& 0.844 $\pm$ 0.076 & 0.878 $\pm$ 0.008 \\
& GCN			& 0.961 $\pm$ 0.005 & 0.962 $\pm$ 0.004 \\
\cdashline{2-4}
& Ours & \textbf{0.983 $\pm$ 0.002} & \textbf{0.982 $\pm$ 0.003} \\
\hline
\multirow{5}{*}{PPI} & DeepWalk 	&0.715 $\pm$ 0.008 &0.706 $\pm$ 0.005\\ 
& node2vec 	&0.773 $\pm$ 0.010 &0.766 $\pm$ 0.005\\ 
& L3			&0.899 $\pm$ 0.003 &0.861 $\pm$ 0.003\\
& VGAE		&0.875 $\pm$ 0.004 &0.844 $\pm$ 0.006\\
& GCN			& 0.894 $\pm$ 0.002 &0.907 $\pm$ 0.006\\
\cdashline{2-4}
& Ours & \textbf{0.907 $\pm$ 0.003} & \textbf{0.918 $\pm$ 0.002} \\
\hline 
\multirow{5}{*}{GDI} & DeepWalk 		& 0.827 $\pm$ 0.007 & 0.832 $\pm$ 0.003  \\
& node2vec 		& 0.828 $\pm$ 0.006 & 0.834 $\pm$ 0.003  \\
& L3			& 0.899 $\pm$ 0.001 & 0.832 $\pm$ 0.001 \\
& VGAE			& 0.902 $\pm$ 0.006 & 0.873 $\pm$ 0.009 \\
& GCN			& 0.909 $\pm$ 0.002 & 0.906 $\pm$ 0.002 \\
\cdashline{2-4}
& Ours  & \textbf{0.933 $\pm$ 0.001} & \textbf{0.945 $\pm$ 0.001} \\
\hline
\end{tabular}
\end{table}

The results on mean AUROC and AUPRC $\pm$ one standard deviation are reported in Table~\ref{table_results}. The mean and standard deviation are calculated across 5 independent splits of positive and negative pairs. Table~\ref{table_results} shows that our proposed inference framework achieves significant improvement over other methods. In comparison to network embedding approaches such as DeepWalk and node2vec, our method gains 22.84\% in AUPRC on the DTI network, 40.83\% on the DDI network, 26.85\% on the PPI network, and 13.09\% on the GDI network over DeepWalk. Although node2vec uses biased random walk and outperforms DeepWalk, our method achieves 19.97\% gain in AUPRC on the DTI network, 22.72\% on the DDI network, 17.34\% on the PPI network, and 12.82\% on the GDI network over node2vec. 

To investigate the contribution of our inference framework to GCN's performance improvement, we compare our method with a fixed-structure GCN encoder with 3 GC layers. The results in Table~\ref{table_results} show that the GC encoder with fixed structure achieves superior performance over both network embedding methods (i.e., DeepWalk and node2vec) and network similarity methods (e.g., L3). Our proposed framework improves the GCN's performance by inferring the most plausible network structure, and achieves 3.35\% improvement in AUPRC on the DTI network, 2.29\% on the DDI network, 1.45\% on the PPI network, and 2.64\% on the GDI network. 

Since the number of layers determines the range of the neighborhood from which the encoder captures information, the performance improvement can be attributed to the fact that our framework allows the GCN aggregates information from the appropriate neighborhood scope by inferring the most plausible depth of the encoder while mitigating over-smoothing.

\subsection*{Calibrating model's prediction}
Given that $p_{ij}$ signifies the probability of interaction, this probability can be interpreted as a measure of the model's confidence in its prediction, where a higher probability indicates greater confidence in a potential interaction. By evaluating the reliability of the model's confidence estimates, we can assess how well-calibrated the model's outputs are. We compare the calibration of our framework with a GCN with a fixed structure.

We adopt Brier score~\cite{brier1950verification} to quantify the reliability of the model's confidence. Brier score is a proper scoring rule for measuring the accuracy of predicted probabilities. A lower Brier score represents better calibration of predicted probabilities. Mathematically, we compute Brier score as the mean squared error of the ground-truth interaction label $A_{ij}$ and predicted probabilities $p_{ij}$.

\begin{equation}
    \text{Brier score} = \frac{1}{|\mathcal{E}|} \sum_{(i,j) \in \mathcal{E}} (A_{ij} - p_{ij})^2 
\end{equation}
where $|\mathcal{E}|$ is the number of edges in the test set.

\begin{table}[htb]
    \centering
    \captionsetup{justification=centering}
    \caption{Brier scores for GCN and our method (lower the better).}
    \label{tab:calibration}
    \begin{tabular}{c|cc}
    \hline
    Method & GCN & Ours \\
    \hline  
    DTI & $0.112\pm0.001$ & $\mathbf{0.109\pm0.002}$\\
    DDI & $0.139\pm0.003$ & $\mathbf{0.042\pm0.001}$\\
    PPI & $0.181\pm0.046$ & $\mathbf{0.119\pm0.002}$\\
    GDI & $0.165\pm0.035$ & $\mathbf{0.111\pm0.001}$\\
    \hline
    \end{tabular}
\end{table}

Table~\ref{tab:calibration} shows the comparison of the Brier score obtained with the predetermined-structure encoder and the one with our inferred structure. Our method gains 2.68\% on the DTI network, 69.78\% on the DDI network, 34.25\% on the PPI network, and 32.73\% on the GDI network. The results indicate that GCN encoder with a predetermined structure makes either over-confident or under-confident predictions. In contrast, by inferring the encoder's network structures and averaging our predictions with respect to all the possibilities, our method achieves well-calibrated predictions with lower Brier scores.

\subsection*{Exploration of drug representations}
\begin{figure}[!htb]
    \includegraphics[width=0.99\linewidth]{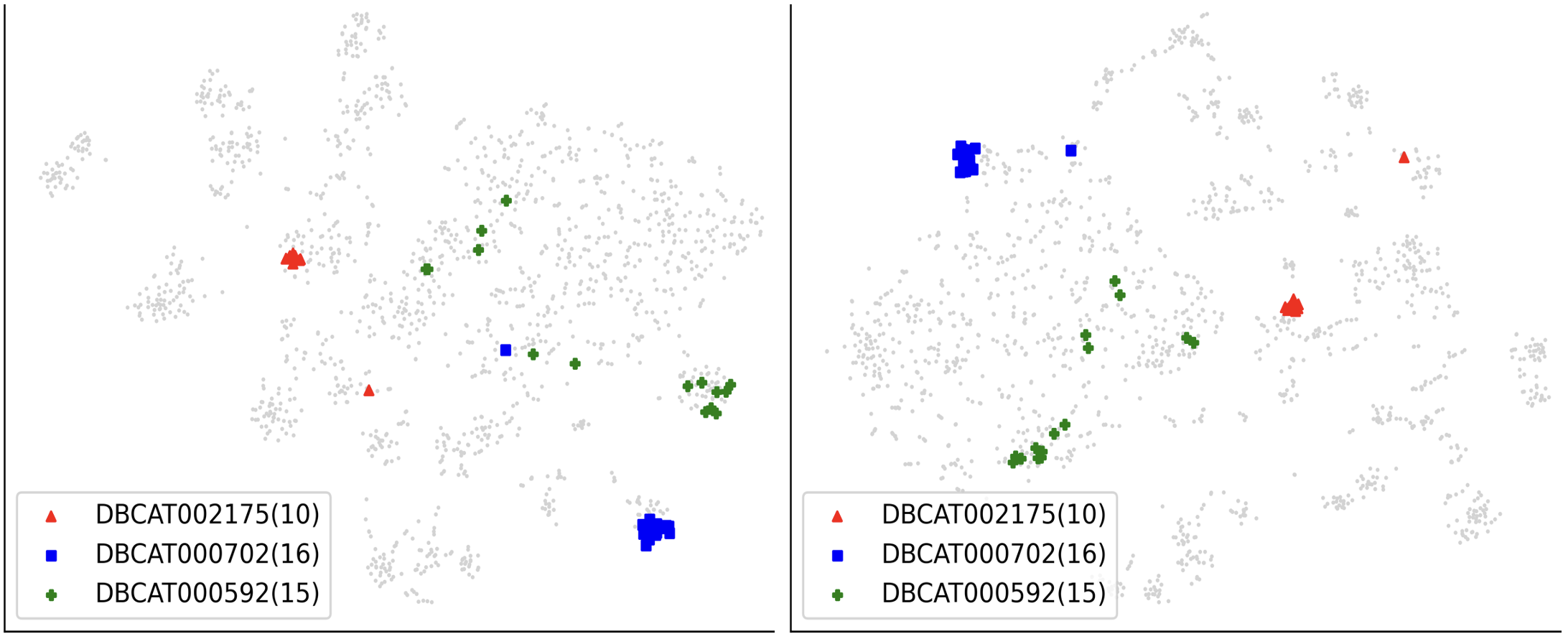}
    \raggedleft \caption{Visualizing the t-SNE embedding of the learned representations for drugs with GCN (left) and our (right) methods. Drugs from three drug categories such as DBCAT002175, DBCAT000702, and DBCAT000592 are highlighted. The numbers in the brackets are the number of drugs from each category.} \label{fig:tsne}
\end{figure}

We assess the quality of the encoded drug representations of the DDI network by visualizing their two-dimensional t-SNE embeddings. The visualization demonstrates how the representations of the drugs with common properties are grouped in the space. We select three categories of drugs from Drugbank \cite{Wishart2006:Drugbank} i.e. ACE Inhibitors and Diuretics (DBCAT002175), Penicillins (DBCAT000702), and Antineoplastic Agents (DBCAT000592). The drugs belonging to these categories are highlighted in Fig. \ref{fig:tsne} with red, blue, and green colors, respectively.

Fig. \ref{fig:tsne} (left) shows that the representations of drugs learned by a GCN are not well separated. Specifically, some of the representations from the Penicillins (blue) and Antineoplastic Agents (green) are close to each other and a clear boundary between the two categories is not established. Furthermore, the representations for the Antineoplastic Agents category are widespread, suggesting that the representation learned by GCN is not able to adequately reflect the similarities among those drugs. In constrast, the representations learned by our method in Fig. \ref{fig:tsne} (right) have well-established cluster boundaries between the three drug categories. Also, the similarities of the drugs from the Antineoplastic Agents category are  well-reflected by their proximity in the representation space.

\begin{table}[htb]
    \centering
    \captionsetup{justification=centering}
    \caption{Novel predictions of GDI based on the evidence from DisGeNET \cite{Pinero2020:Disgenet}.}
    \label{tab:novel_gdi}
    \begin{tabular}{cccc}
    \hline
    Gene & Disease & GCN & Ours \\
    \hline
    ADD2 & Hypertensive Disease & 0.30 & 0.89 \\
    ADD2 & Lupus Erythematosus & 0.29 & 0.71 \\
    HLA-E & Nasopharyngeal Carcinoma & 0.46 & 0.64 \\
    SERPINA3 & Alzheimer's Disease & 0.38 & 0.57 \\
    GSDME & Melanoma & 0.57 & 0.96 \\
    GSDME & Malignant Neoplasm (Lung) & 0.49 & 0.94 \\
    GSDME & Malignant Neoplasm (Stomach) & 0.58 & 0.95 \\
    \hline
    \end{tabular}
\end{table}

\begin{table}[htb]
    \centering
    \captionsetup{justification=centering}
    \caption{Novel predictions of DDI with the evidence supporting the interactions.}
    \label{tab:novel_ddi}
    \begin{tabular}{ccccc}
    \hline
    Drug & Drug & GCN & Ours & Evidence \\
    \hline
    Nelfinavir & Acenocoumarol & 0.95 & 0.99 & \cite{Garcia1999:Sequential} \\
    Praziquantel & Itraconazole & 0.95 & 0.99 & \cite{Perucca2006:Clinically} \\
    Dapsone & Warfarin & 0.38 & 0.63 & \cite{Truong2012:Probable} \\
    \hline
    \end{tabular}
\end{table}

\subsection*{Investigation of novel predictions}
We evaluate the confidence of the predictions by comparing the predictive probabilities of GCN and our method on the true interactions that are in the up-to-date databases. Yet, they are not included in the training datasets that are the GDI and DDI networks collected in 2018. Table \ref{tab:novel_gdi} shows novel gene-disease pairs with the predictive probabilities from GCN and our method. The evidence of these novel interactions is provided in DisGeNET \cite{Pinero2020:Disgenet}. Our method predicts the existence of these interactions with a 35\% higher probability on average. GCN misses five out of seven interactions, given that a probability of higher than 50\% is considered as positive predictions. On the contrary, our method consistently captures the new interactions. Similarly, the novel predictions for drug-drug pairs in Table \ref{tab:novel_ddi} show our method's superior predictive ability compared to GCN. These results suggests that probabilistic inference on the GCN encoder structure leads to more confident predictions on the true interactions.

\subsection*{Performance robustness against network sparsity}
We examine the robustness of our proposed framework against network sparsity and compare its performance with a fixed-structure GCN encoder. We train the model on the varying percentage of training edges from 10\% to 70\%. We consider 20\% of the interaction dataset as the test set to evaluate the predictive performance.  

\begin{figure*}[!htb]
\centering
\subfigure{\raggedleft\label{dti_train_ratio}\includegraphics[width=0.24\linewidth]{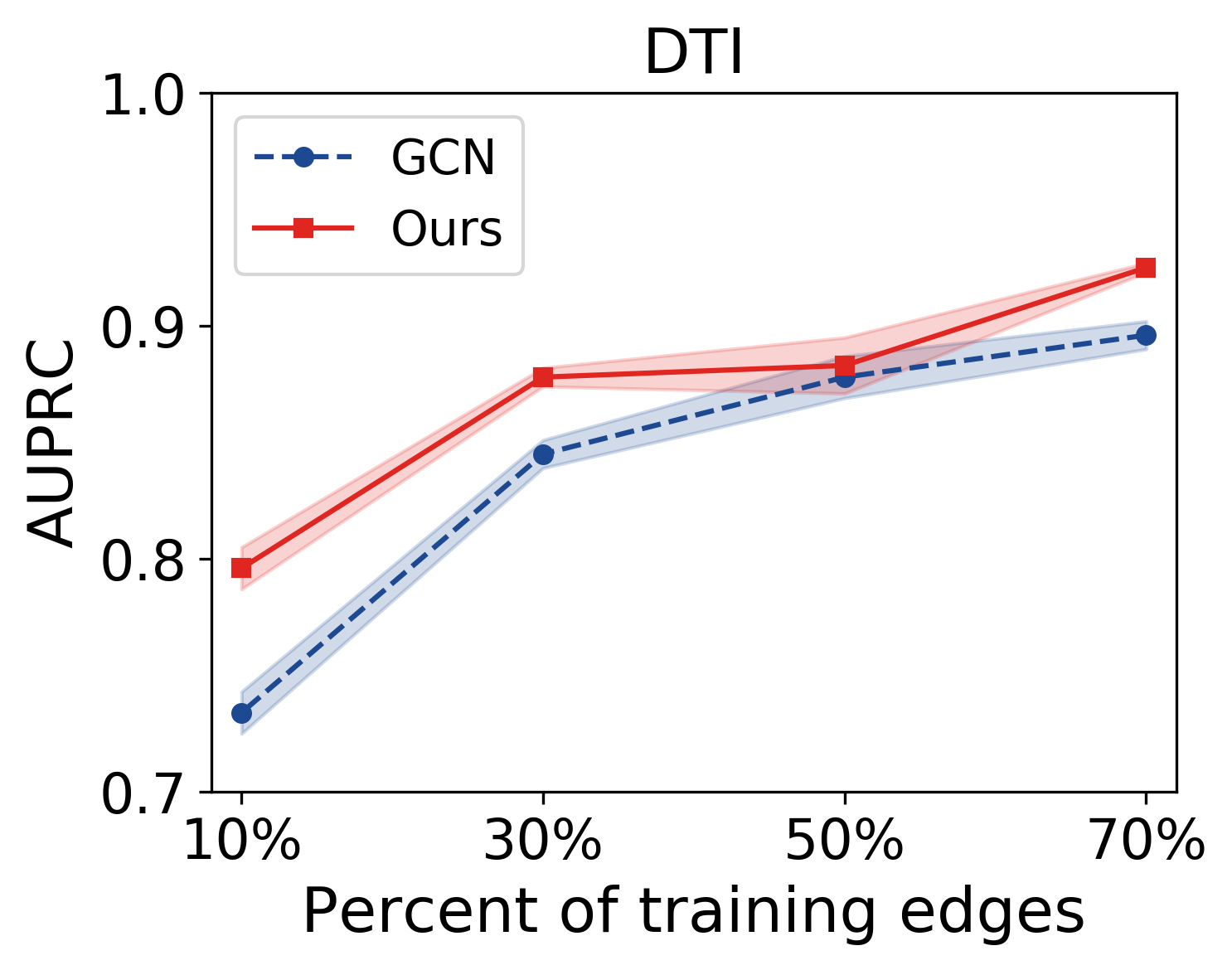}}
\subfigure{\raggedleft\label{ddi_train_ratio}\includegraphics[width=0.24\linewidth]{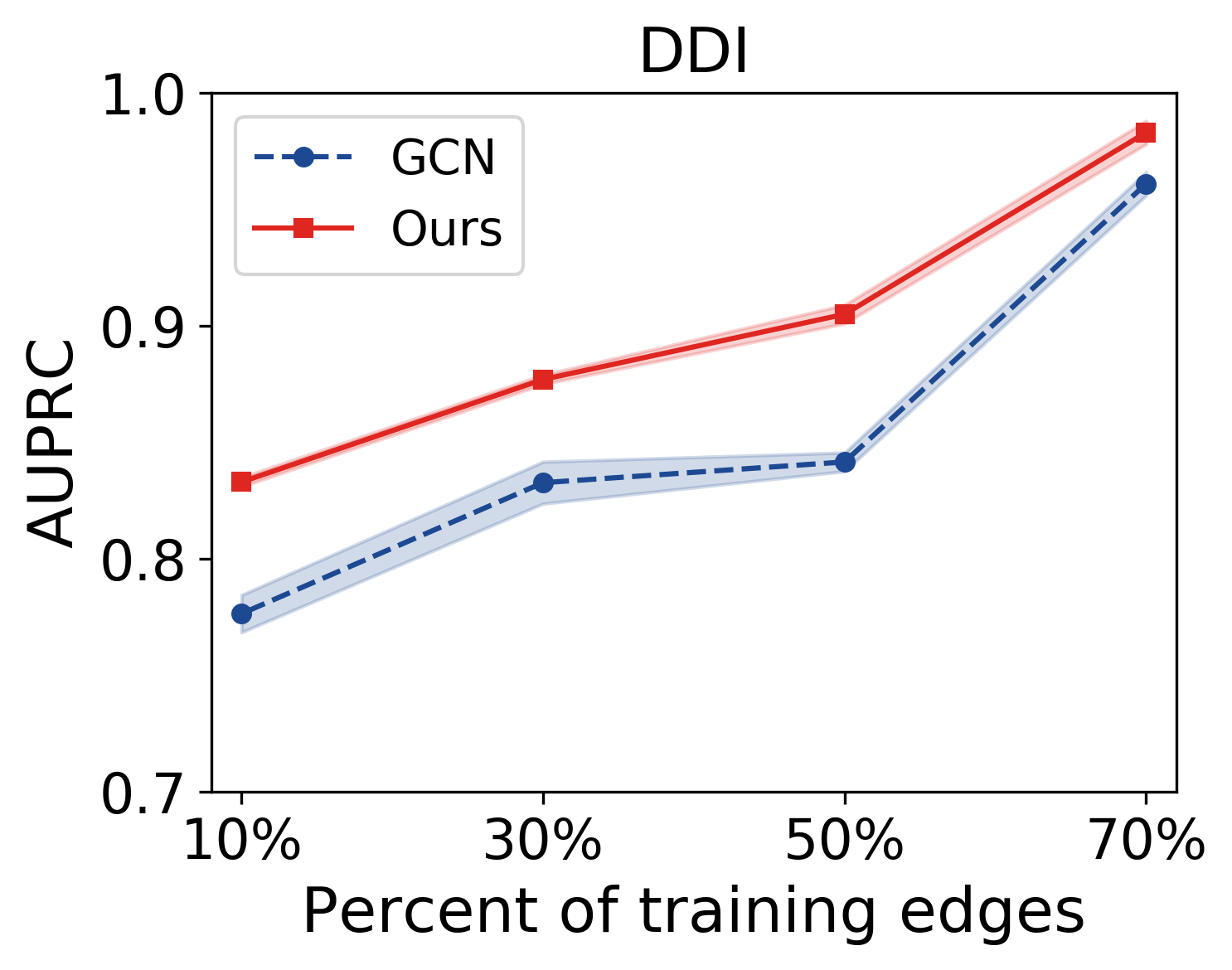}}
\subfigure{\raggedleft \label{ppi_train_ratio}\includegraphics[width=0.24\linewidth]{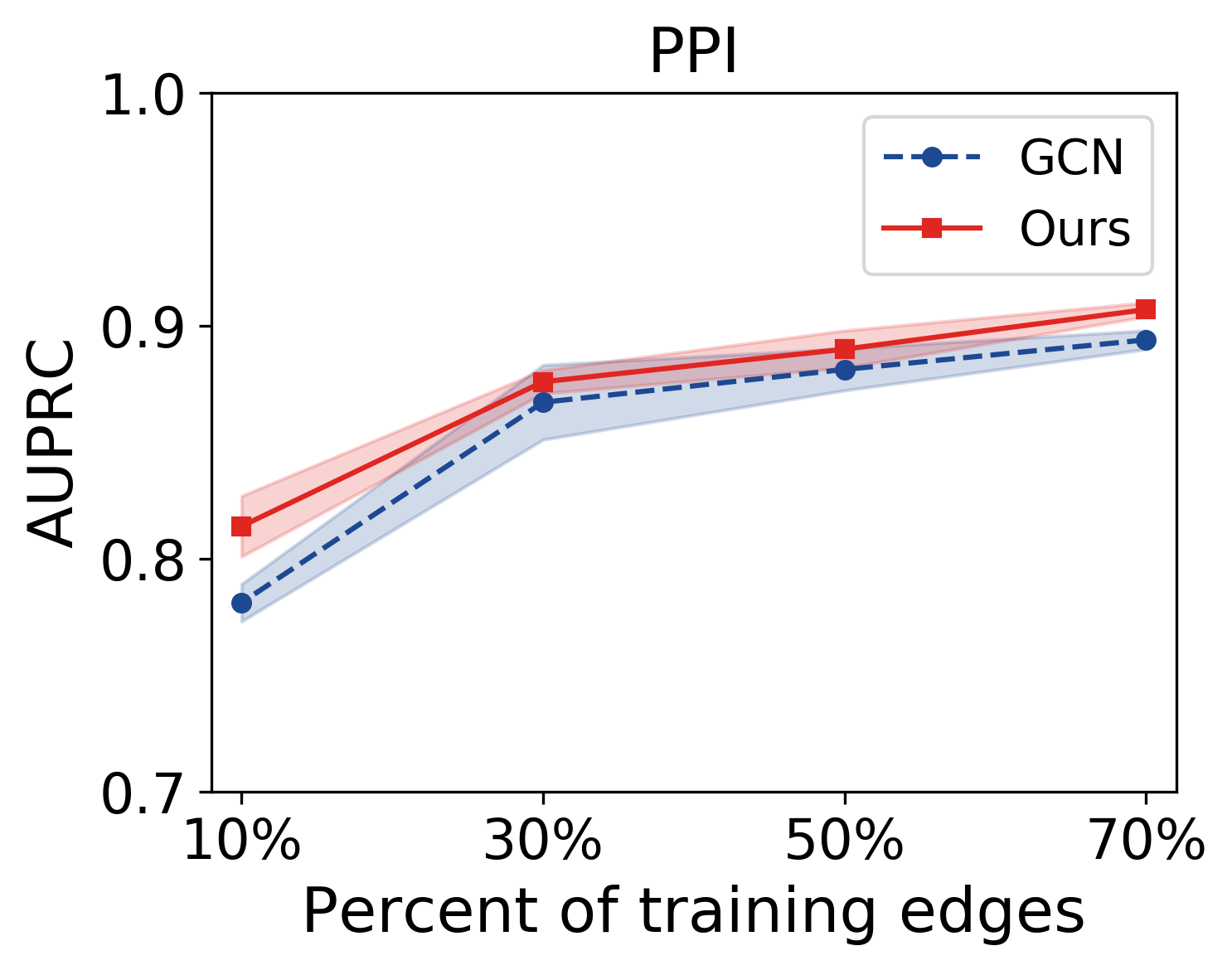}}
\subfigure{\raggedleft \label{gdi_train_ratio}\includegraphics[width=0.24\linewidth]{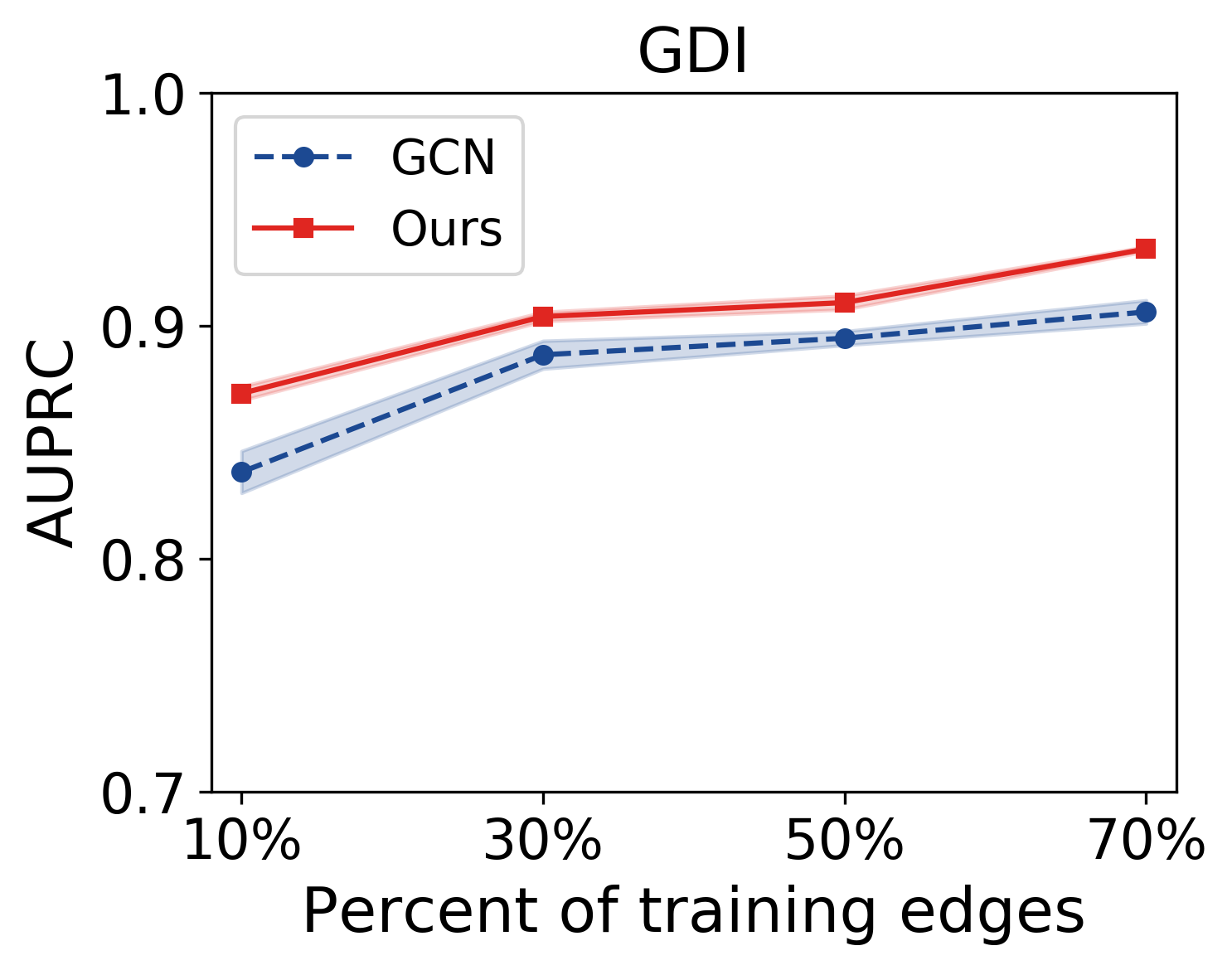}}
\raggedleft \caption{AUPRC comparison for our method and GCNs when trained with different fractions of training edges. (from left to right) DTI, DDI, PPI, and GDI.} \label{train_ratio}
\end{figure*}

\begin{figure*}[!htb]
    \centering
     \subfigure{\raggedleft\label{dti_sparsity_ratio}\includegraphics[width=0.24\linewidth]{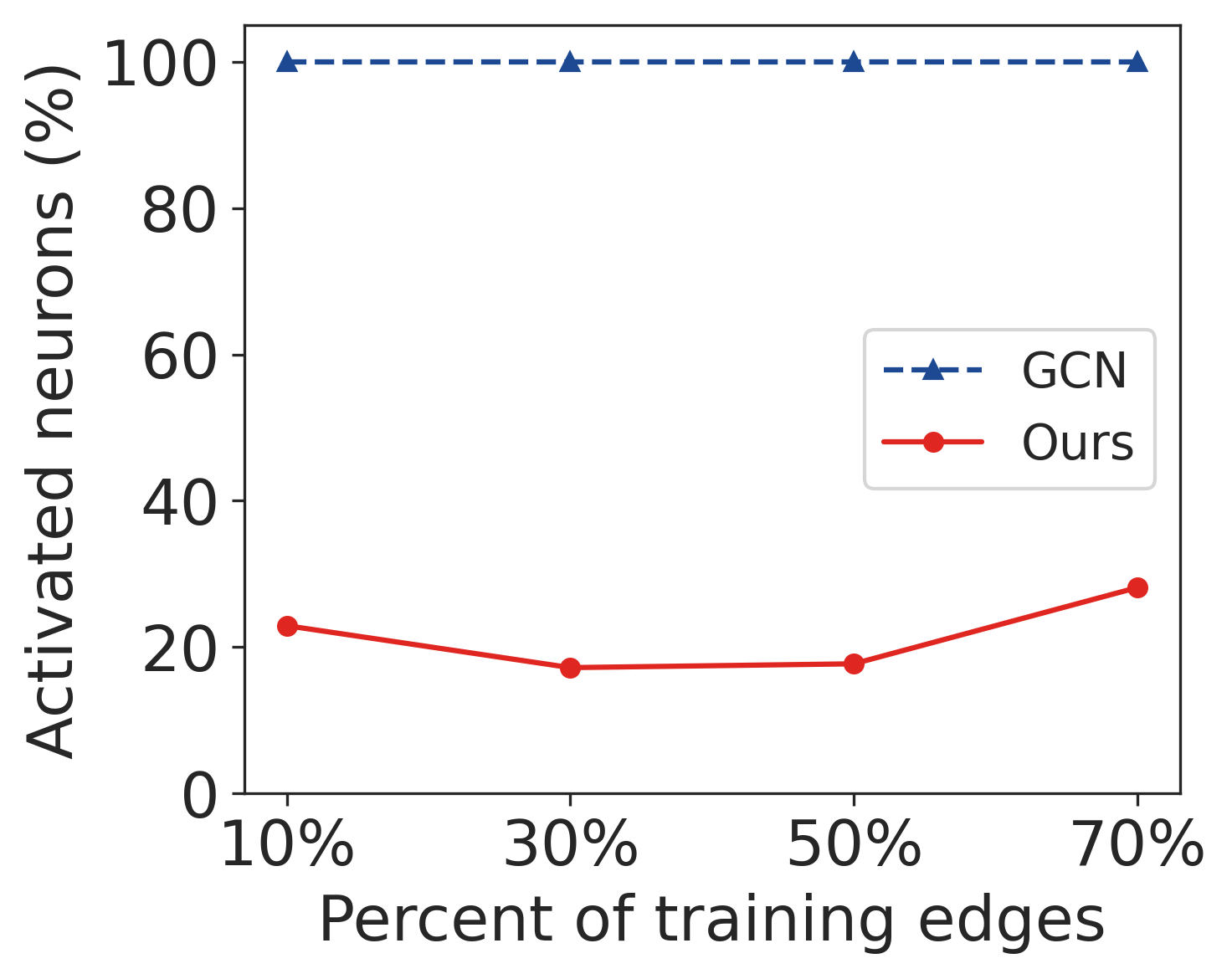}}
     \subfigure{\raggedleft\label{ddi_sparsity_ratio}\includegraphics[width=0.24\linewidth]{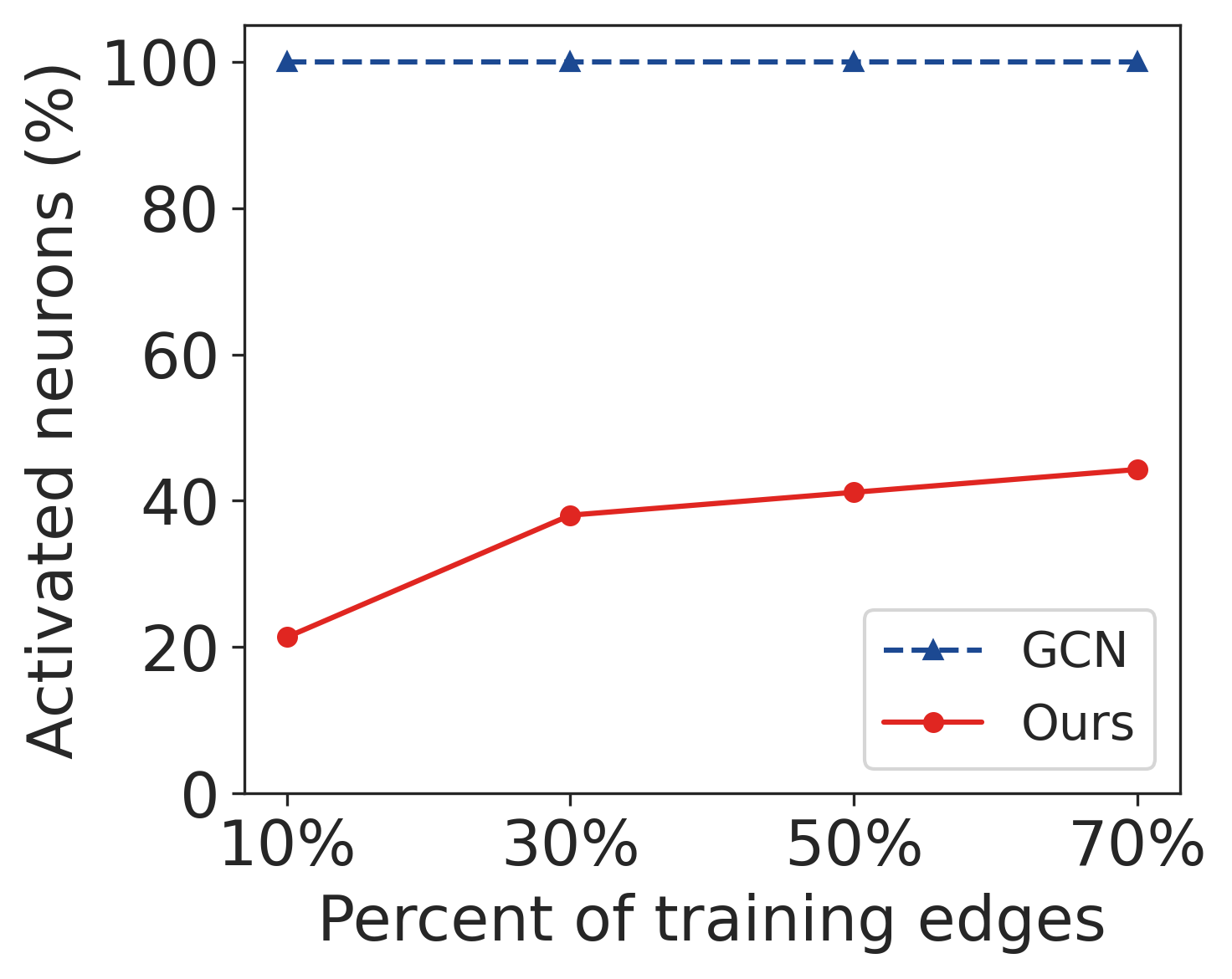}}
     \subfigure{\raggedleft \label{ppi_sparsity_ratio}\includegraphics[width=0.24\linewidth]{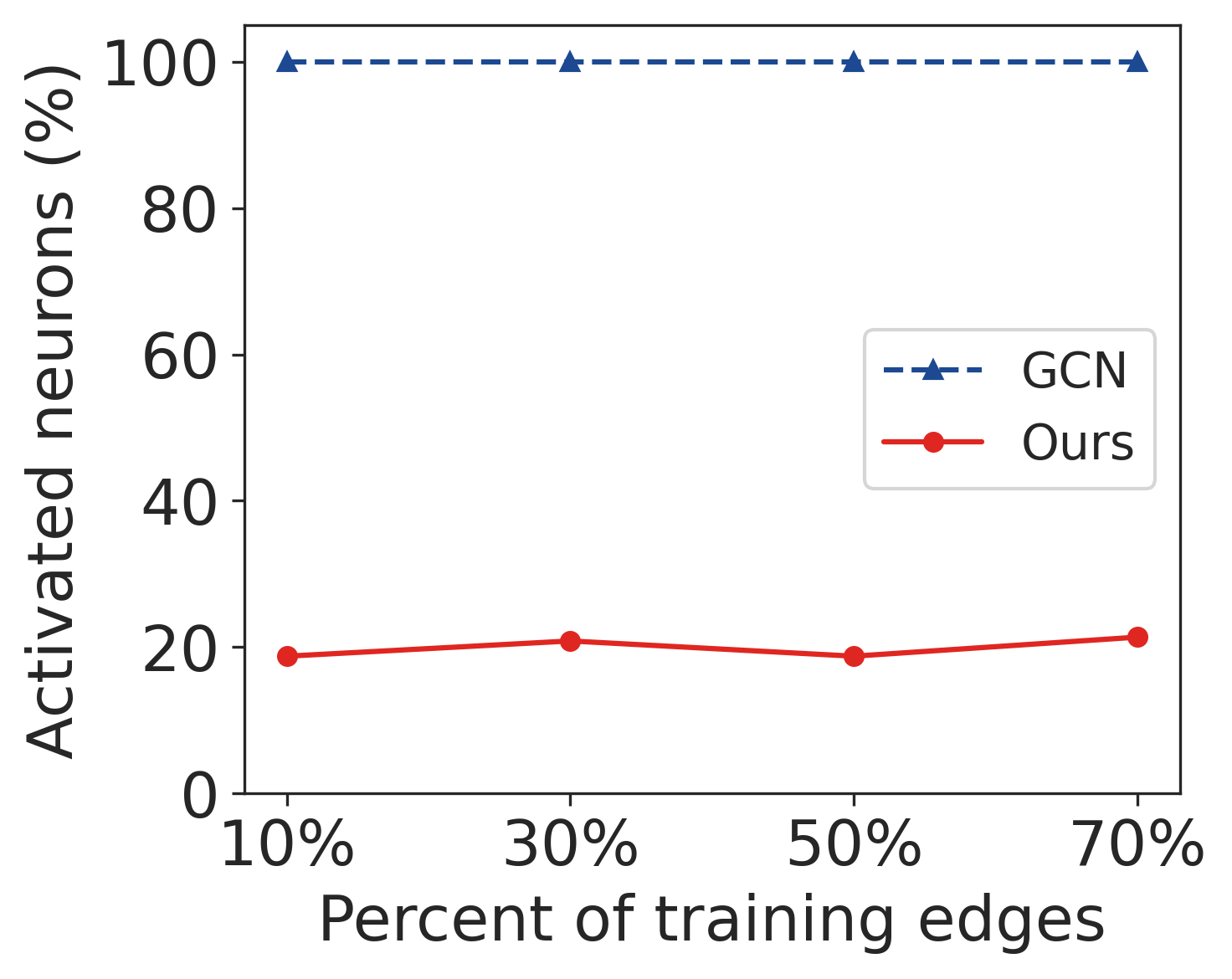}}
     \subfigure{\raggedleft \label{gdi_sparsity_ratio}\includegraphics[width=0.24\linewidth]{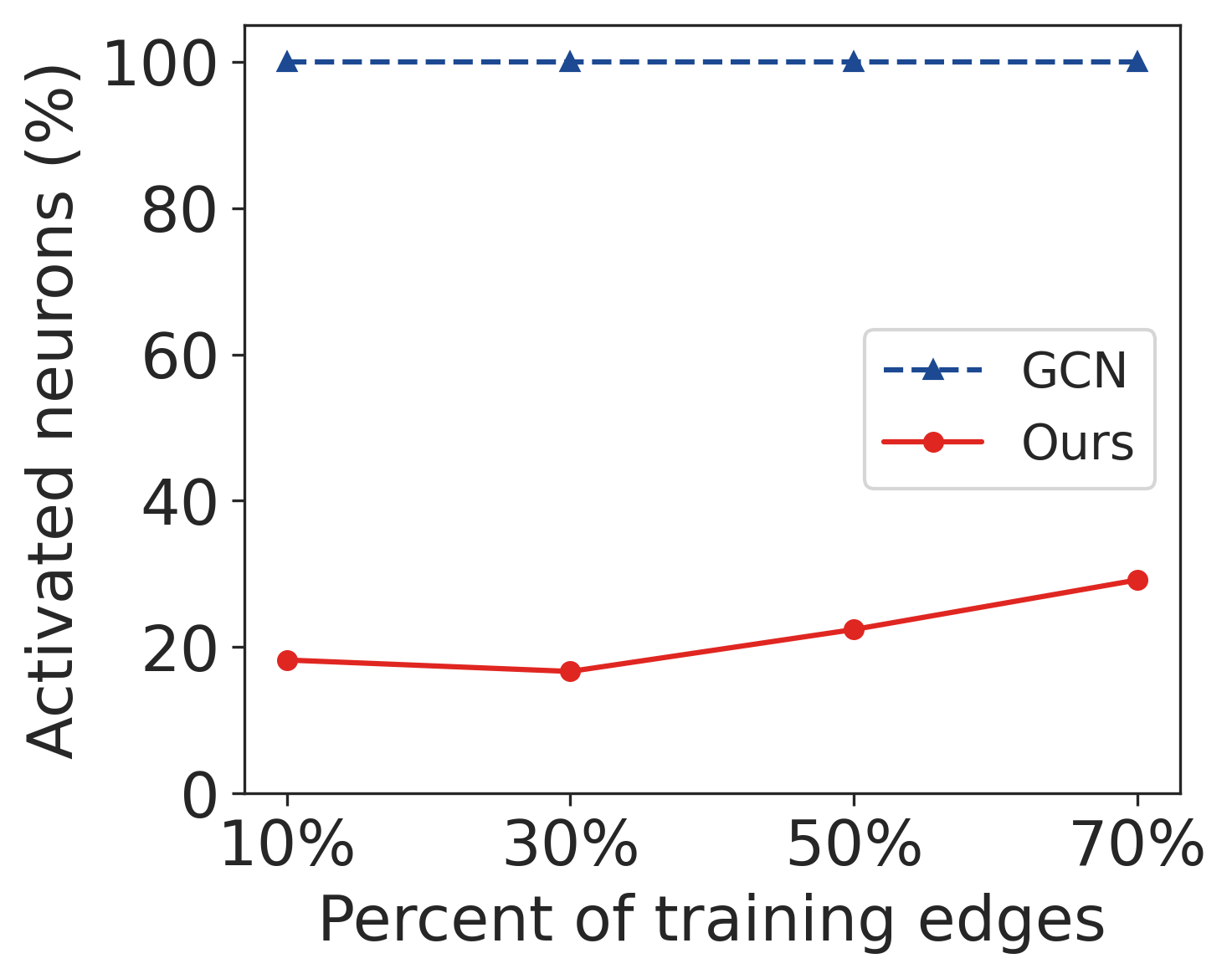}}
    \raggedleft \caption{Comparison of activated neurons for our method and GCNs when trained with different number of training interactions. (from left to right) DTI, DDI, PPI and GDI.} \label{sparsity_ratio}
\end{figure*}

Figure~\ref{train_ratio} shows that our proposed method is better at maintaining high performance for biomedical networks with high sparsity levels, indicating it is more robust to network sparsity. To gain further insights into the robustness, we examine the trend of the percentage of activated neurons in our model as network sparsity increases. Figure~\ref{sparsity_ratio} illustrates that the percentage activation of the neurons increases with an increasing number of training edges (decreasing sparsity). In contrast, the fixed-structure activates all its neurons across different sparsity levels of the training datasets. This suggests that the robustness of our model to data sparsity is gained from the capability of our model that dynamically adapts its structure in response to the data sparsity to achieve optimal performance.

\subsection*{Effect of truncation level $K$}
We next investigate the influence of the depth ($L$) and the truncation level ($K$) on the performance of the fixed-structure GCN encoder model and our model, respectively. 

\begin{figure*}[!ht]
    \centering
     \subfigure{\raggedleft\label{dti_layers}\includegraphics[width=0.24\linewidth]{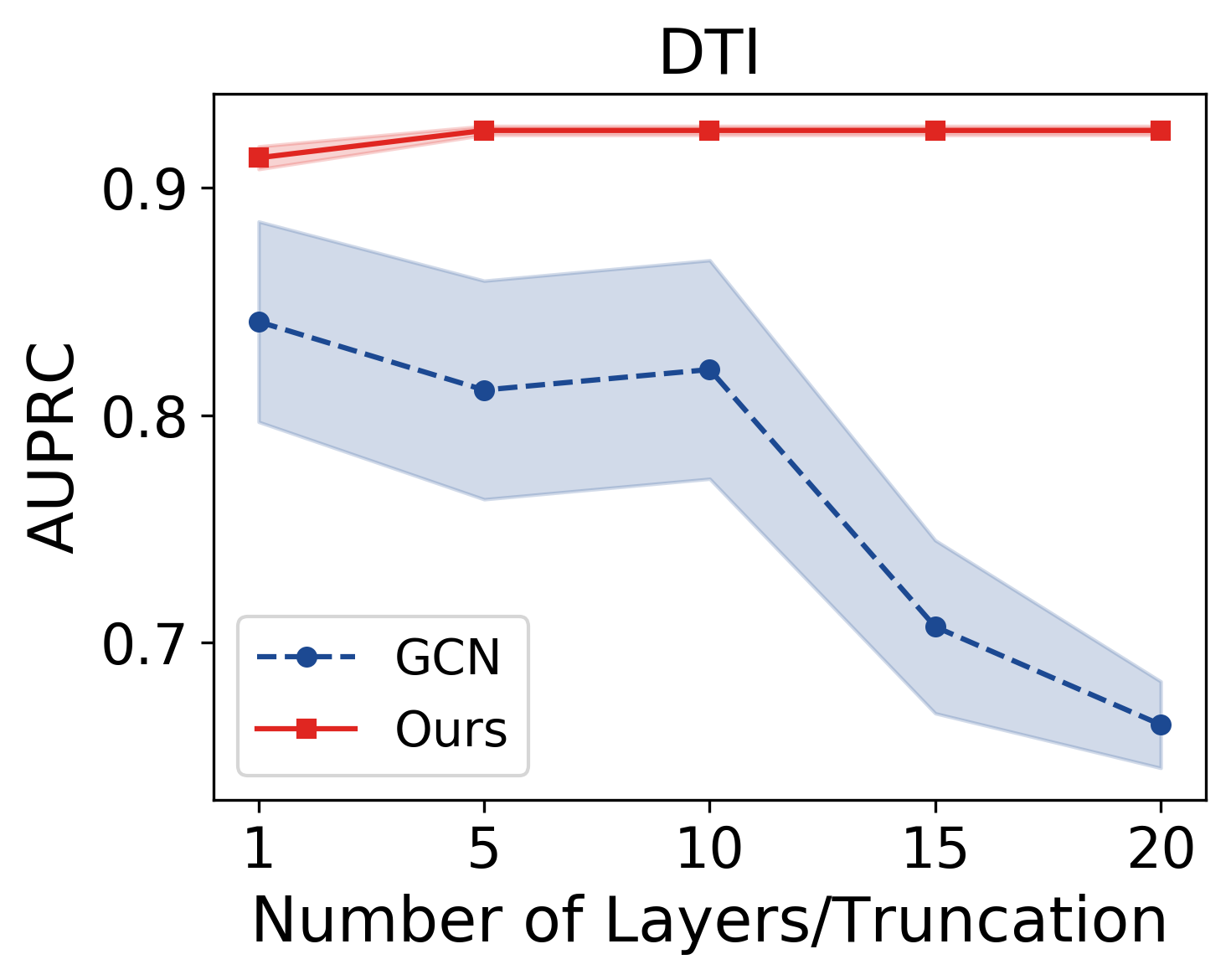}}
     \subfigure{\raggedleft\label{ddi_layers}\includegraphics[width=0.24\linewidth]{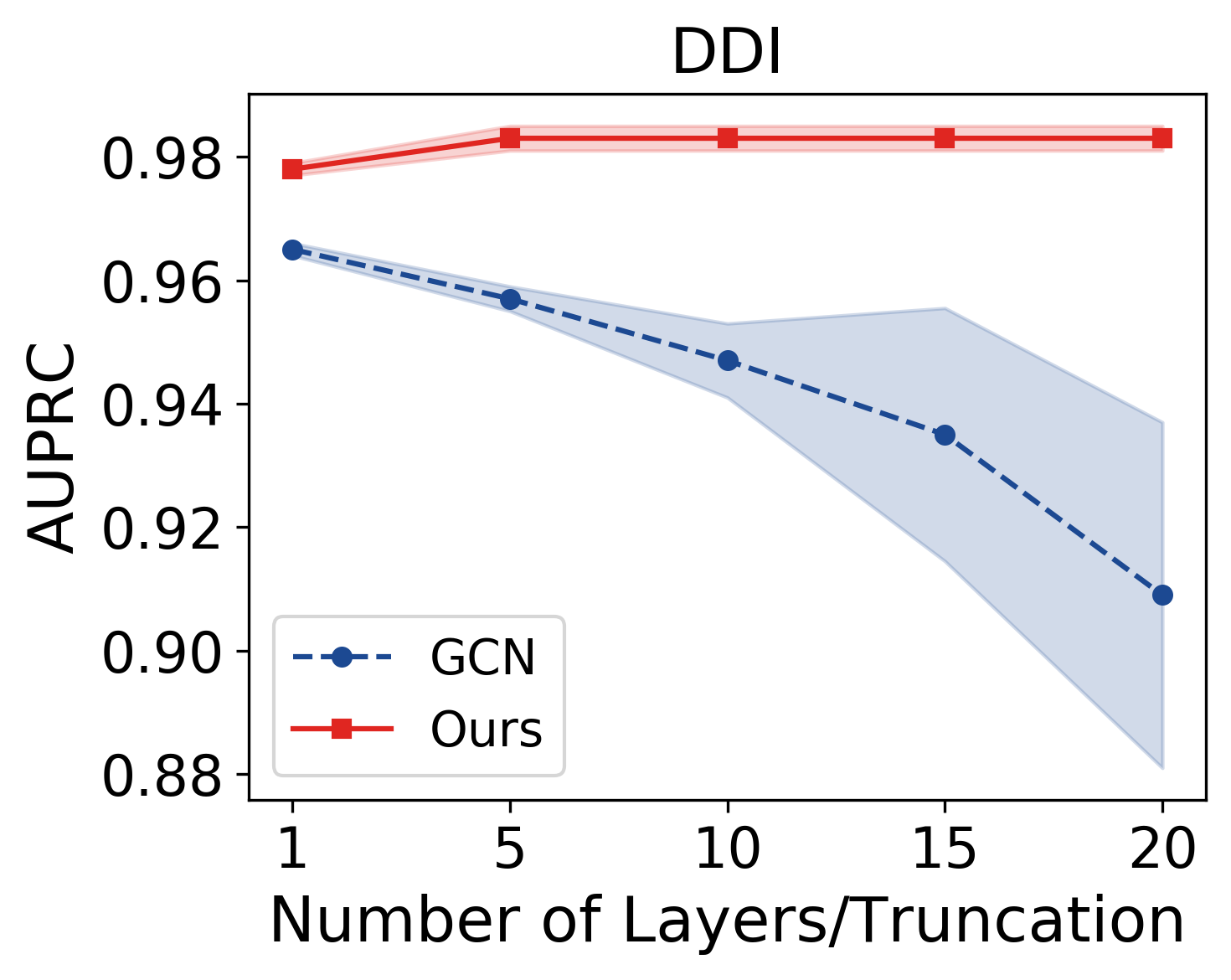}}
     \subfigure{\raggedleft \label{ppi_layers}\includegraphics[width=0.24\linewidth]{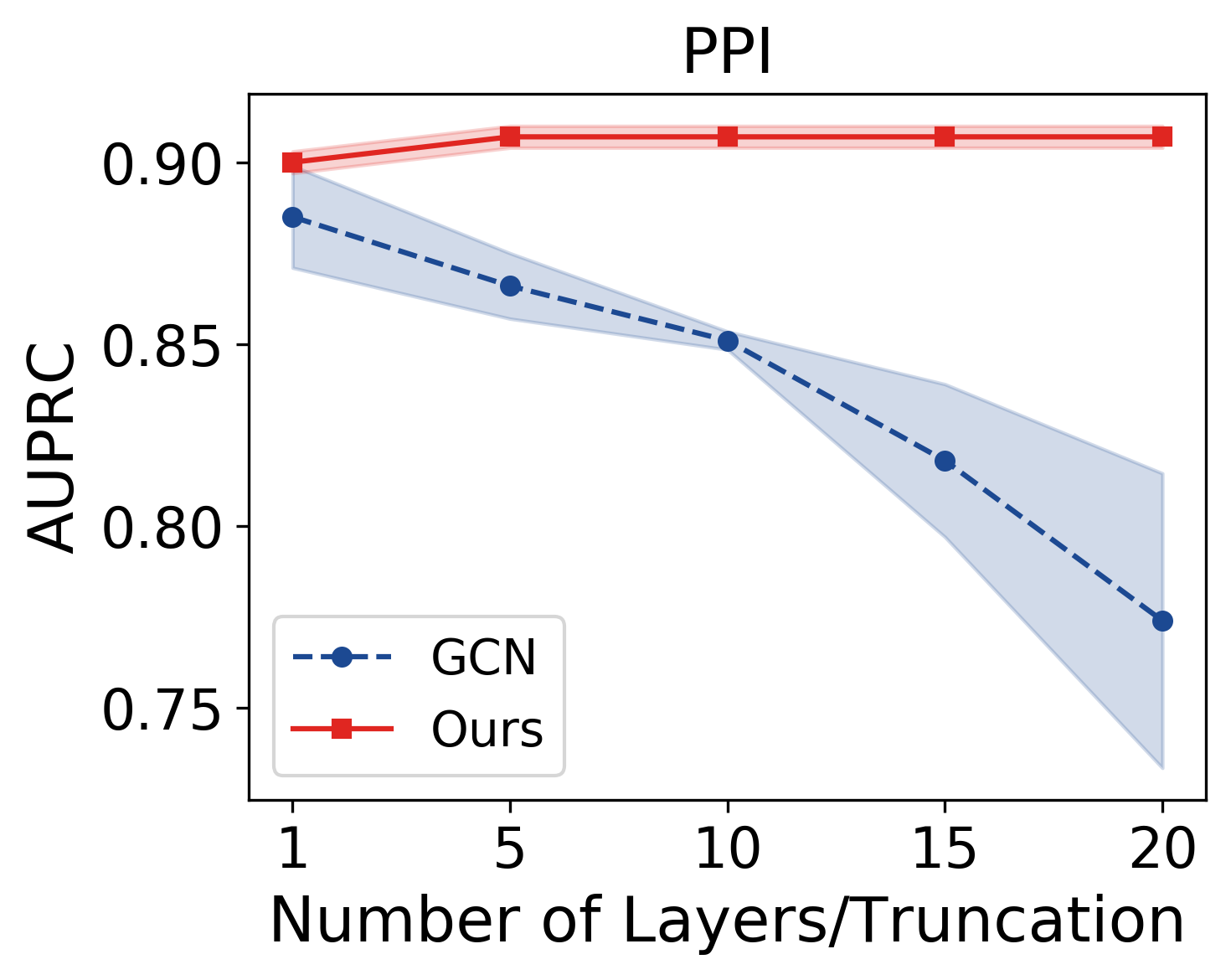}}
     \subfigure{\raggedleft \label{gdi_layers}\includegraphics[width=0.24\linewidth]{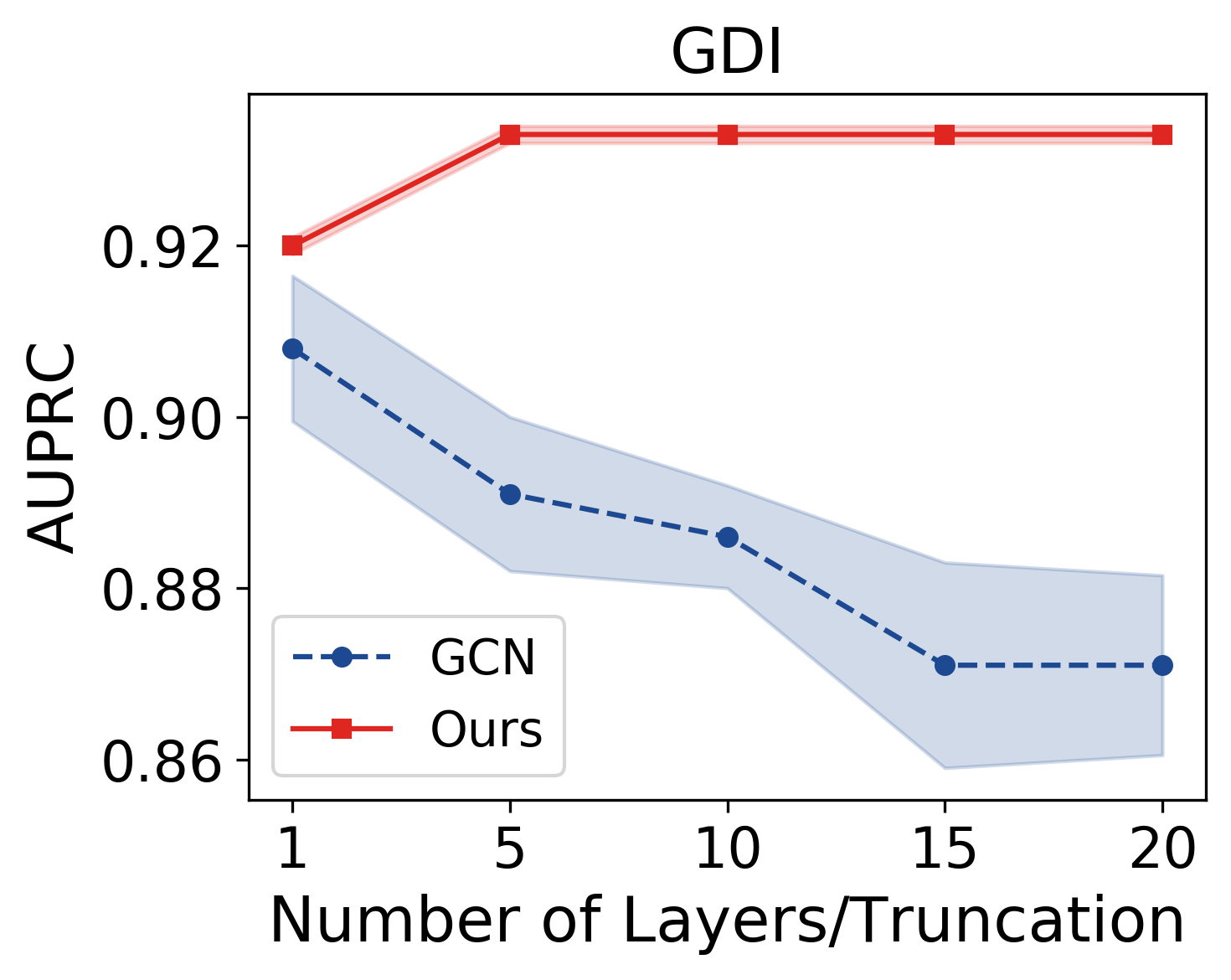}}
    \raggedleft \caption{AUPRC comparison for our method with different truncation and GCNs with different number of layers.  (from left to right) DTI, DDI, PPI and GDI.}
    \label{depth_truncation}
\end{figure*}

The baselines we compare with are fixed-structure GCN encoders with different depths $L$ over the range $L \in \{1, 5, 10, 15, 20\}$. Similarly, we train our proposed framework with truncation level $K$ over the same range $K \in \{1, 5, 10, 15, 20\}$. For both methods, we set the maximum number of neurons $M$ in each hidden layer in the encoder to 64. Figure~\ref{depth_truncation} shows that although our method has an underfitting problem at $K=1$ due to limited model capacity, it achieves overall better performance. When the truncation level $K$ becomes sufficiently large $(K\geq 5)$, the performance becomes independent of $K$. The results are consistent with Bayesian model selection theorem in~\cite{kc2021joint}. 

Figure~\ref{depth_truncation} shows that the GC encoder with predetermined structures suffers from an overfitting problem with the increase in the depth of the encoder. Shallow models with a single hidden layer perform better compared to their deeper counterparts. The result suggests that our method is robust to overfitting by jointly inferring the encoder depth and their neuron activations. Essentially, our method balances the encoder depth and their network activations to achieve better performances.

\begin{figure*}[!htb]
    \centering
    \subfigure{\raggedleft\label{dti_sparsity}\includegraphics[width=0.24\linewidth]{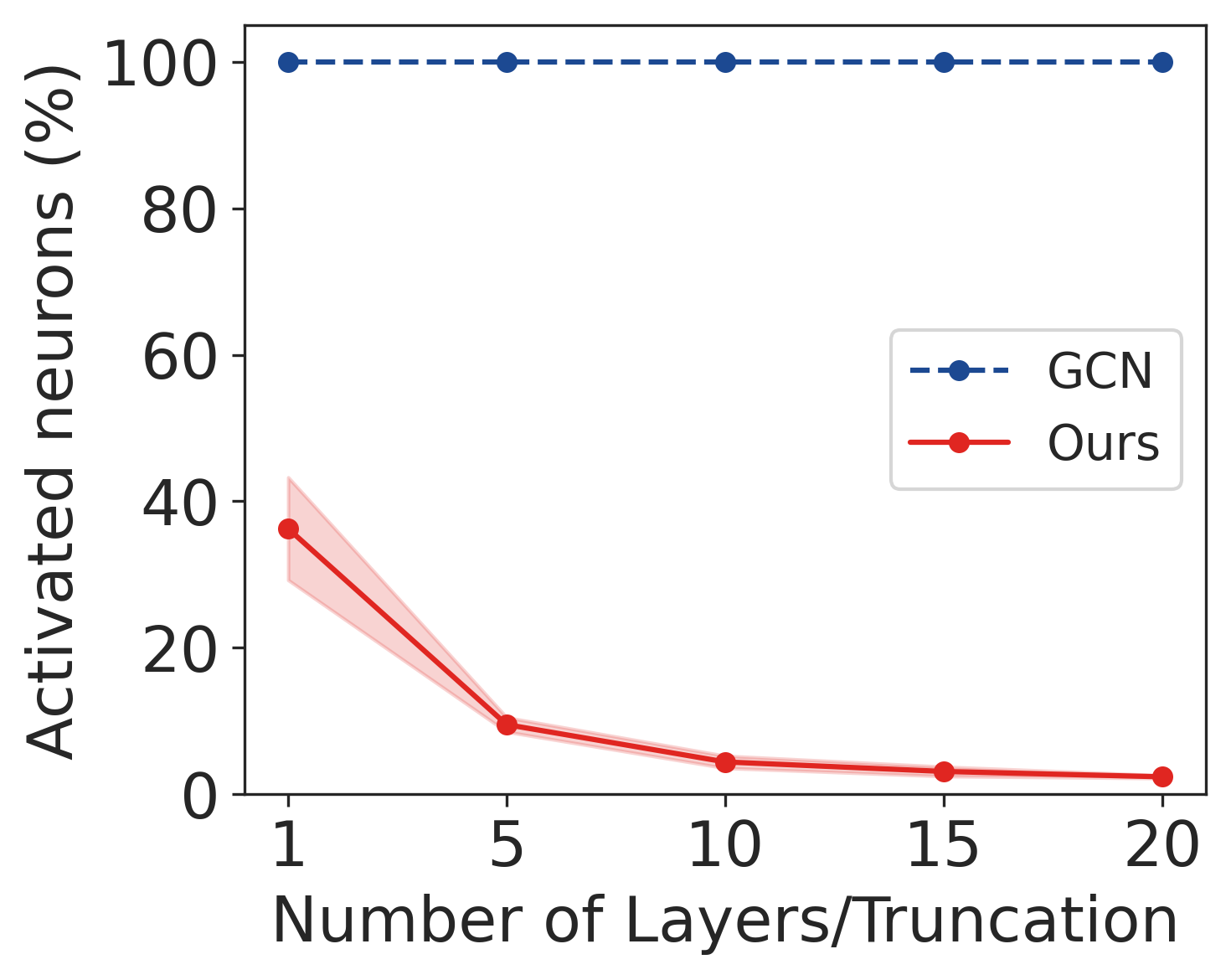}}
    \subfigure{\raggedleft\label{ddi_sparsity}\includegraphics[width=0.24\linewidth]{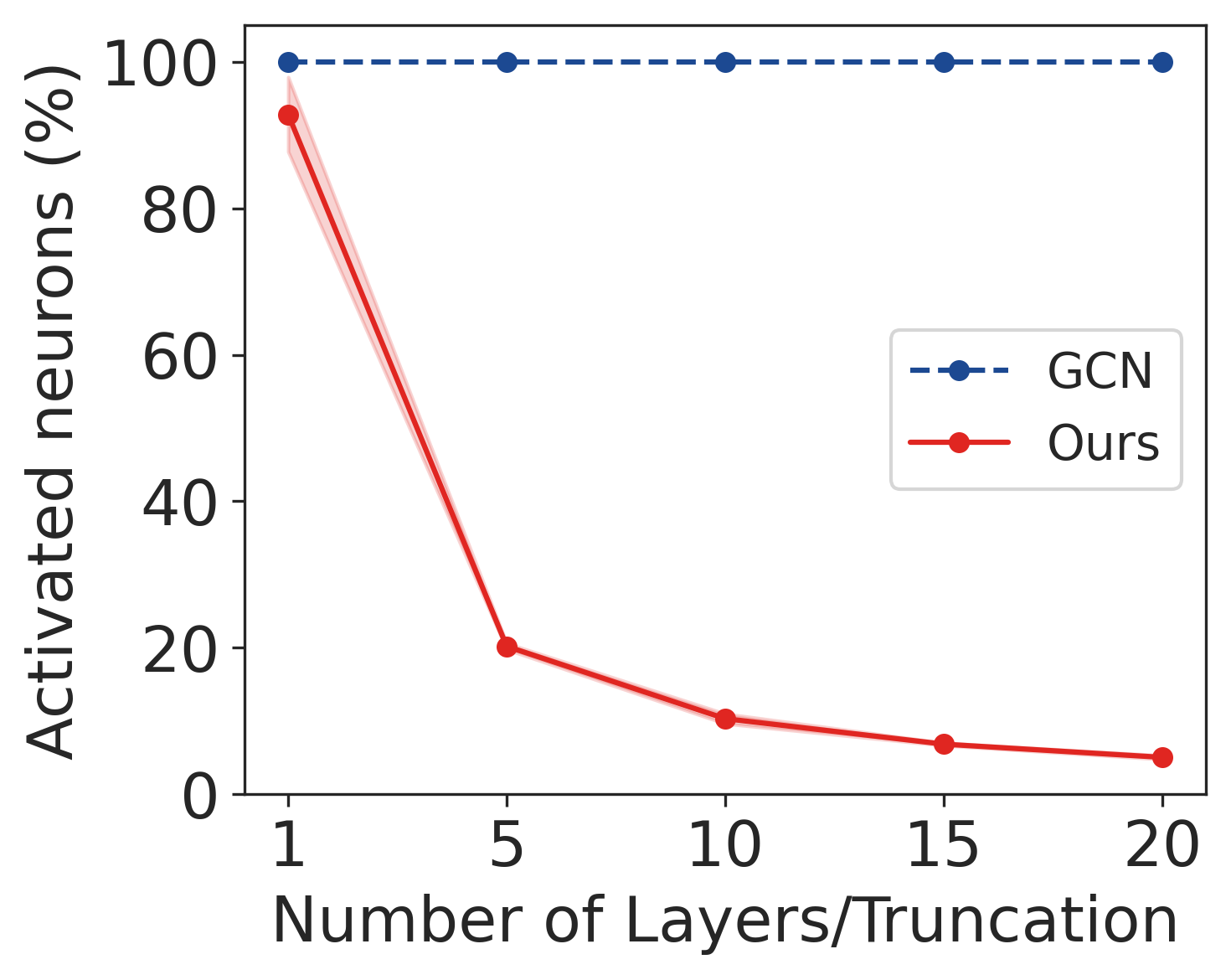}}
     \subfigure{\raggedleft \label{ppi_sparsity}\includegraphics[width=0.24\linewidth]{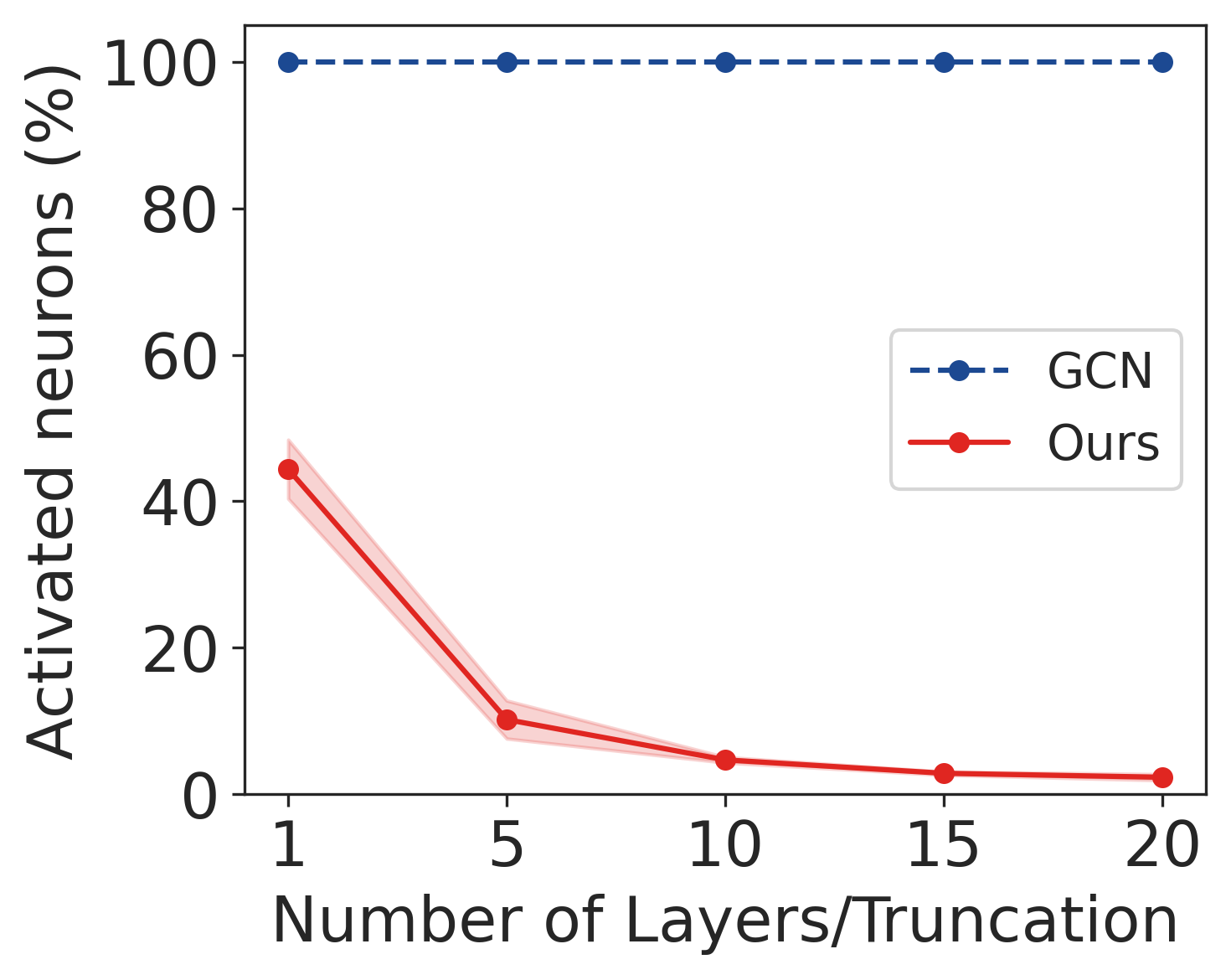}}
     \subfigure{\raggedleft \label{gdi_sparsity}\includegraphics[width=0.24\linewidth]{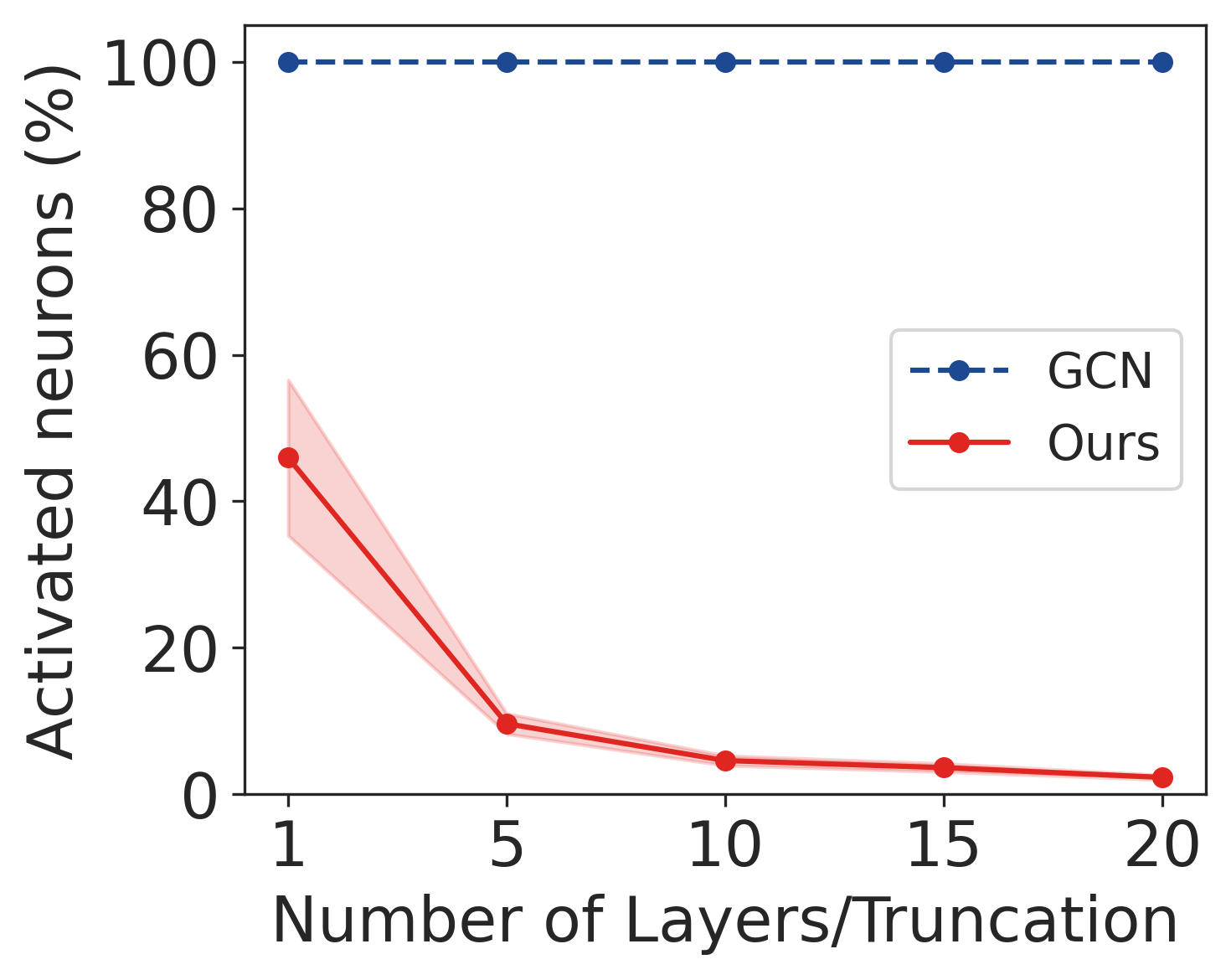}}
    \raggedleft \caption{Comparison of activated neurons for our method and GCNs when trained with different truncation $K$ or depth $L$. (from left to right) DTI, DDI, PPI and GDI.} \label{network_structure}
\end{figure*}

Figure~\ref{network_structure} shows the percentage of activated neurons with respect to truncation $K$ and depth $L$, respectivly. For GCN models with fixed encoder structure, all neurons are activated and thus have an overfitting problem. On the contrary, our proposed method automatically regularizes the model complexity by activating a relatively small number of neurons. With the settings $K=20$, our model activates only 5\% neurons.

\section{Conclusion}
In this paper, we present a Bayesian inference framework to infer the neural network structures of the GC encoders. We demonstrate that our method enables the encoder structures to dynamically evolve to accommodate the interactions that are incrementally available over time. The experiments for biomedical interaction prediction show that we achieve accurate and well-calibrated predictions with compact encoder network structures and significantly fewer model parameters. Additionally, compared to GCN, our method learns representations that more accurately reflect drug relationships and makes correct and stronger predictions of novel interactions, which are subsequently validated by the literature.

\section*{Acknowledgments}
This work was supported by the NSF [NSF-1062422 to A.H.] and [NSF-1850492 and NSF-2045804 to R.L.].





\bibliographystyle{ieeetr}
\bibliography{reference}

\end{document}